\documentclass[12pt,twoside]{report}
\usepackage{english,a4}
\usepackage{epsf}
\usepackage{epsfig}
\usepackage{times}

\begin{document}

\vspace*{0.6cm}

\begin{quote}

\begin{center}
{\Large 
Static and dynamical properties of a supercooled liquid
confined in a pore\footnote{Talk presented in the workshop {\it Dynamics in
Confinement} Grenoble, 26 - 29 January, 2000}}

Peter Scheidler, Walter Kob, and Kurt Binder

{\small \it 
Institut f\"ur Physik, Johannes Gutenberg-Universit\"at Mainz,
Staudinger Weg 7, D-55099 Mainz, Germany }
\end{center}
\end{quote}

\vspace{0.7cm}

\hspace*{1mm} \hfill
\begin{minipage}[t]{165mm}
{\footnotesize
{\bf Abstract}. We present the results of a Molecular Dynamics computer
simulation of a binary Lennard-Jones liquid confined in a narrow pore.
The surface of the pore has an amorphous structure similar to that of
the confined liquid. We find that the static properties of the liquid are
not affected by the confinement, while the dynamics changes dramatically.
By investigating the time and temperature dependence of the intermediate
scattering function we show that the dynamics of the particles close
to the center of the tube is similar to the one in the bulk, whereas
the characteristic relaxation time $\tau_q(T,\rho)$ of the intermediate
scattering function at wavevector $q$ and distance $\rho$ from the axis
of the pore increases continuously when approaching the wall, leading
to an apparent divergence in the vicinity of the wall. This effect is
seen for intermediate temperatures down to temperatures close to the
glass transition.  The $\rho$-dependence of $\tau_q(T,\rho)$ can be
described by an empirical law of the form $\tau_q(T,\rho)=f_q(T) \exp
[\Delta_q/(\rho_p-\rho)]$, where $\Delta_q$ and $\rho_q$ are constants,
and $f_q(T)$ is the only parameter which shows a significant temperature
dependence. }
\end{minipage}

\vspace{0.75cm}

\noindent {\bf 1. INTRODUCTION}
\vspace{12pt}

\noindent
The dynamics of a bulk liquid in its supercooled state has been
investigated extensively in experiments and computer simulations
and is understood reasonably well [1,2].  Much less is known about
the influence of a spatial confinement on the dynamic properties of
a liquid.  The growing interest in this topic in recent years is based
on the fact that new nanoscale materials with adjustable pore size, such
as Vycor glass, have been developed, which allow to study the influence
of the confinement. Concerning the details of the glass transition, experiments
have so far given controversial results. Depending on the nature of the
pores and the contained liquid, some authors report an increase in the
glass temperature~[3,4], while others find a decrease~[5,6].  In this
paper we report the results of computer simulations which were done to
investigate this phenomenon. Within such simulations (see also~[7,8]) it
is possible to control the nature of the wall (roughness and wall-liquid
interaction), and to do a local analysis of the dynamics of the particles.
Therefore this method is well suited to increase our understanding of
the effects of the confinement.
\vspace{0.75cm}

\noindent {\bf 2. MODEL AND DETAILS OF THE SIMULATION}
\vspace{12pt}

\noindent 
To mimic the experimental setup of a fluid confined in porous materials
we take  as the spatial confinement a cylindrical tube.  The contained
liquid is chosen to be a simple Lennard-Jones fluid. To prevent
crystallization at low temperatures we take a binary mixture of 80\% A
and 20\% B particles with the same mass and interacting via a Lennard-Jones
potential of the form $V_{\alpha\beta}(r)=4\epsilon_{\alpha\beta}
[(\sigma_{\alpha\beta}/r)^{12}-(\sigma_{\alpha\beta} /r)^{6}]$ with
$\alpha,\beta \in\{{\rm A,B}\}$ and cut-off radii $r_{\alpha,\beta}^C$=$2.5
\cdot \sigma_{\alpha\beta}$.  The parameters were chosen as
$\epsilon_{\rm AA}$=1.0, $\sigma_{\rm AA}$=1.0, $\epsilon_{\rm AB}$=1.5,
$\sigma_{\rm AB}$=0.8, $\epsilon_{\rm BB}$=0.5, and $\sigma_{\rm BB}$=0.88.  
The bulk
properties of this system have been investigated in the past~[9].  In the
following, all results will be given in reduced units, i.e. length in
units of $\sigma_{\rm AA}$, energy in units of $\epsilon_{\rm AA}$ and time
in units of $(m\sigma_{\rm AA}^2/48\epsilon_{\rm AA})^{1/2}$. For Argon these
units correspond to a length of 3.4\AA, an energy of 120K$k_B$ and a
time of $3\cdot10^{-13}$s.

To minimize the influence of the changes in {\it static} properties
due to the confinement, such as layering or a change in the static
structure factor, on changes in its {\it dynamic} properties, we chose
the wall of the pore to have an amorphous structure similar to the one
of the confined liquid.  For this purpose, we equilibrated a large
bulk system at an intermediate temperature, $T$=0.8, and extracted
a cylinder with radius $\rho_T+r_{AA}^C$ with a tube radius $\rho_T$
of 5.0.  During the simulation of the tube the particles in the outer
ring, $\rho$$\ge$$\rho_T$, remained fixed while the inner particles,
$\rho$$<$$\rho_T$, interact with each other and the wall particles and
were allowed to move.  (Here $\rho$ is the distance from the center of
the cylinder.)

The time evolution of the system was calculated by solving the equations
of motion with the velocity form of the Verlet algorithm with a time
step of 0.01 at high ($T$$\ge$1.0) and 0.02 at low ($T$$\le$0.8)
temperatures.  To improve the statistics we simulated between 8 and
16 independent systems, each containing 1905 fluid particles and about
2300 wall particles.  The tube length of 20.137 was chosen such that the
average particles density is 1.2, the same value as used in the earlier
simulations  of the bulk.  The temperatures investigated were $T$=2.0,
1.0, 0.8, 0.7, 0.6, and 0.55.  The equilibration was done by periodically
coupling the liquid to a stochastic heat bath.  All data presented here
was produced during a microcanonical run at constant energy and volume.

\vspace{0.75cm}

\noindent {\bf 3. RESULTS}
\vspace{12pt}

\begin{figure}[t]
\begin{center}\begin{picture}(2000,286)
        \epsfxsize=180mm\put(0,0){\epsfbox{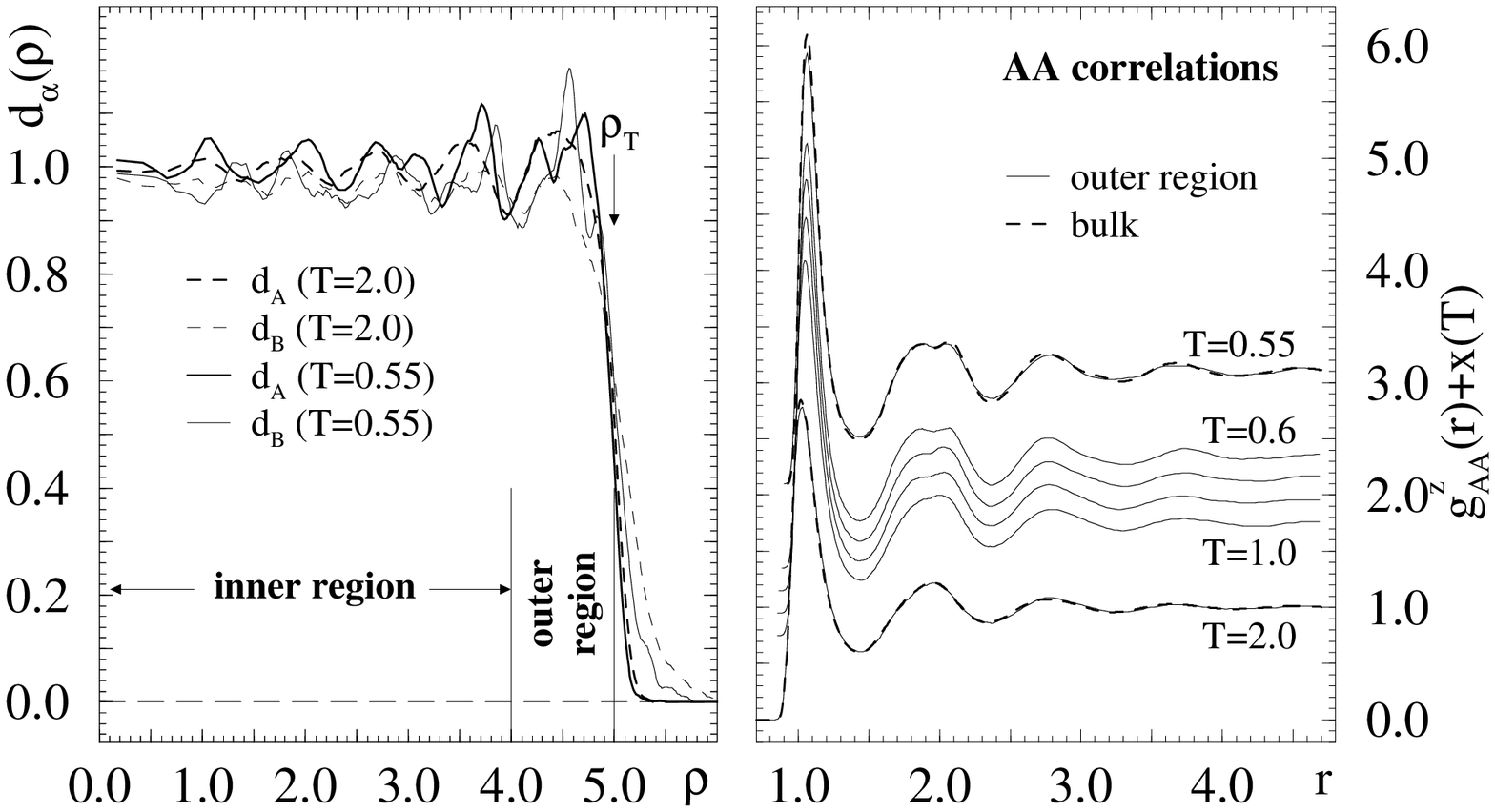}}
\end{picture}
\parbox{170mm}{
{\label{fig12}
\parbox[t]{7.8cm}{
%\begin{flushleft}
\footnotesize {\bf Figure~1.~}Density profiles for A and B particles at $T$=2.0
and $T$=0.55.
%\end{flushleft}
}
\hspace{4mm}
\parbox[t]{8.4cm}{
%\begin{flushleft}
\footnotesize {\bf Figure~2.~}Radial distribution function $g^z_{\rm AA}(r)$ 
plus vertical offset $x(T)$ in the outer region for $T$=2.0 ($x$=0.0),
$T$=1.0 ($x$=0.75), $T$=0.8 ($x$=0.95), $T$=0.7 ($x$=1.15), 
$T$=0.6 ($x$=1.35), and $T$=0.55 ($x$=2.1); 
comparison with bulk curves for $T$=2.0 and $T$=0.55. \vspace{4mm}
%\end{flushleft}
}
}}
\end{center}
\end{figure}
\noindent 
The analysis of the static properties of the confined system gave the 
expected results. 
Looking at the density profile in a plane perpendicular to the 
axis of the tube, normalized to its average value,
\begin{equation}
d_{\alpha}(\rho)=\left \langle \int_{interior}{\sum_{i=1}^{N_{\alpha}}
{\delta \left(\sqrt{x_i^2+y_i^2} - \rho \right)}} \right \rangle \cdot 
\left[ \frac{N_{\alpha}}{V} \right]^{-1} \mbox{, where } \alpha \in 
\{\rm A,B\},
\label{eq_profile}
\end{equation}
we observe only a small dependence of $d$ on the distance $\rho$ from the
center (Fig.~1).  Due to the roughness of the surface the fluid particles
can penetrate slightly into the wall.  We can define a penetration radius
$\rho_p$ as the distance from the center of the tube at which there
is almost no chance of finding a particle, namely the value of $\rho$
were the density profile has decreased to $10^{-4}$ from its bulk value.
For A particles we find $\rho_A$$\approx$5.5$\pm$0.2, and for the smaller
B particles $\rho_A$$\approx$6.1$\pm$0.2, values that depend only weakly
on temperature below $T$=0.7.  From the figure we see that the fluid has
a tendency to form concentric layers, especially at low temperatures,
but that this effect is only weak.

The radial distribution function in $z$-direction,
\begin{equation}
g_{\alpha \beta}^z(r) \propto \left \langle {\sum_{i=1}^{N_{\alpha}}
\sum_{j=1 \atop x_{ij}^2+y_{ij}^2 < \mu^2}^{N_{\beta}} 
{\delta \left( r - \left| \vec r_i - \vec r_j \right| \right)}
} \right \rangle , \,\, \alpha \in \{\rm A,B\} \,\, , \,\, \mu^2=0.5^2   
\label{eq_raddis}
\end{equation}
shows no strong deviation from its bulk behavior (Fig.~2). Even at the lowest 
temperature, $T$=0.55, there is only a small difference in 
peak positions and amplitudes between the bulk curve and the one for particles in
the outer region, $\rho$$\ge$4.0, of the tube. At high $T$ this difference can 
hardly be seen. Similar results are obtained for AB- and BB-correlations 
and also the static structure factor is hardly affected by the confinement.

In contrast to this, the dynamical properties of the system change
dramatically due to the confinement. All investigated dynamic quantities
(mean squared displacement, intermediate scattering function, and van
Hove correlation function) show a strong $\rho$-dependence. The following
results were obtained by labeling a particle with its distance from
the $z$-axis at time $t$=0, and analysing its dynamics as a function of
its position.  Fig.~3 shows the $\rho$-dependence of the self part of the
intermediate scattering function, 
\begin{equation} F_s(q,\rho,t)=\langle
\exp \left[ i {\vec q} \cdot \left({\vec r}(t)-{\vec r} \right) \right]
\cdot \delta \left( x^2(0)+y^2(0) -\rho^2 \right)
\rangle , \label{eq_fq} \end{equation} 
for A particles and $\vec q$ along the $z$-axis with modulus $q$ corresponding
to the maximum of the static structure factor for AA correlations at the lowest
temperature investigated ($T$=0.55).
\begin{figure}[t]
\begin{center}\begin{picture}(2000,279)
        \epsfxsize=180mm\put(0,0){\epsfbox{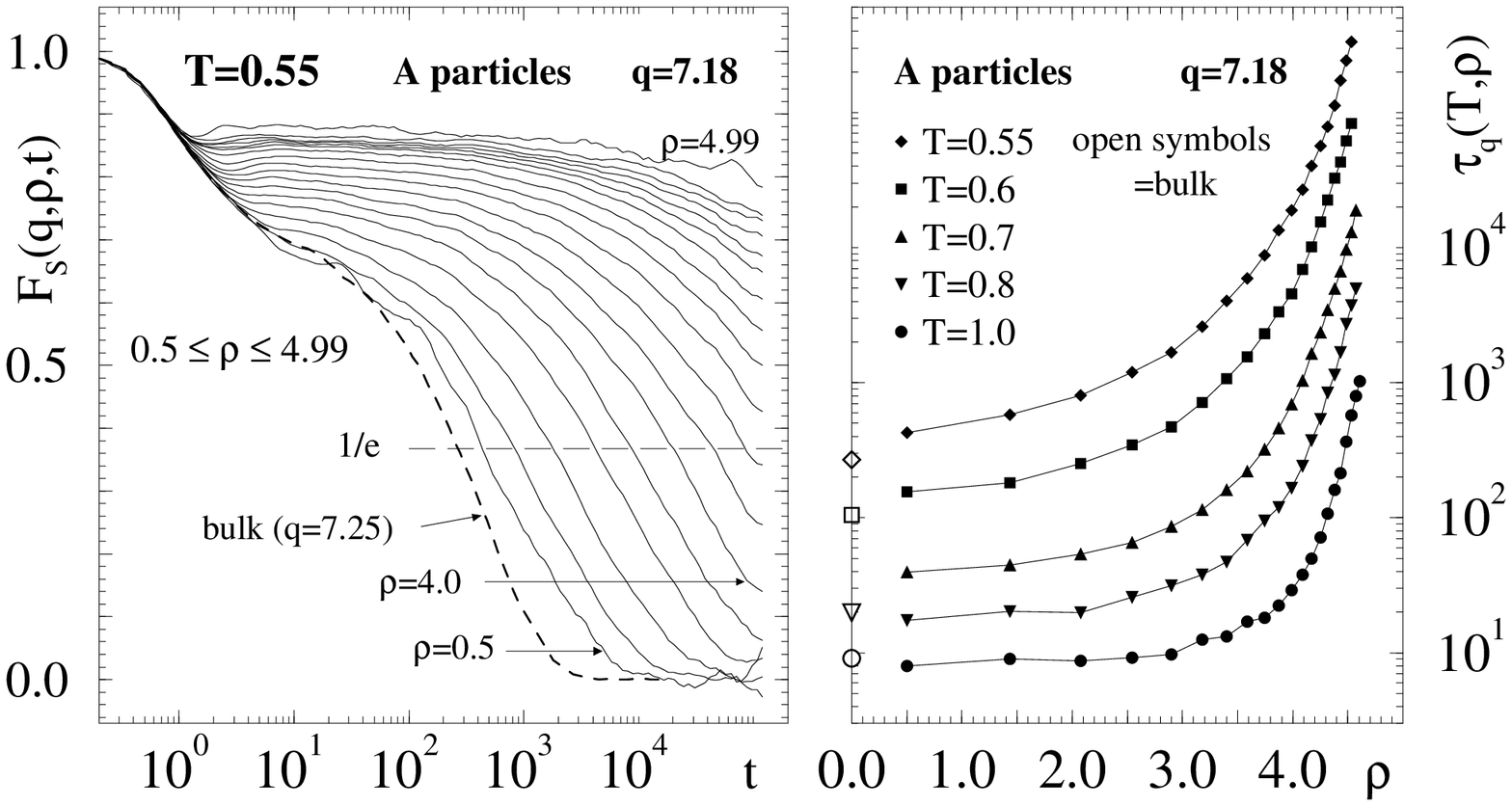}}
\end{picture}
\parbox{170mm}{
{\label{fig34}
\parbox[t]{9cm}{
%\begin{flushleft}
\footnotesize {\bf Figure~3.~}Time dependence of the self part of the 
intermediate scattering function for different values of $\rho$ for A 
particles at $T$=0.55.
%\end{flushleft}
}
\hspace{5mm}
\parbox[t]{7cm}{
%\begin{flushleft}
\footnotesize {\bf Figure~4.~}$\rho$-dependence of the relaxation times 
$\tau_q(T,\rho)$ of $F_s(q,\rho,t)$, compared with the corresponding bulk 
values.\vspace{4mm}
%\end{flushleft}
}
}}
\end{center}
\end{figure}
While the relaxation of $F_s(q,\rho,t)$ in the center is similar to the
one in the bulk, it becomes much slower with increasing $\rho$, i.e.~on
approach to the wall.  The slowing down of the dynamics when approaching
the wall exceeds more than three orders of magnitude. Since any particle
tagged at time zero moves within a range of radii with different intrinsic
relaxation times during the run, the averaged relaxation is more stretched
than in the bulk, especially close to the wall, where the differences
between relaxation times is large.

In order to investigate the $\rho$-dependence of the dynamics we define
a characteristic $\rho$-dependent relaxation time $\tau_q(T,\rho)$
of the intermediate scattering function as the time at which it has
decayed to $e^{-1}$ of its initial value and compare these times.
A quantitative analysis of the $\rho$-dependence of this relaxation
times for the investigated temperatures below $T$=1.0 give the following
results (Fig.~4). At higher temperatures particles in the inner region,
$\rho$$\le$2.0, show almost bulk behavior before a strong increase in
$\tau_q(T,\rho)$ becomes apparent for higher $\rho$. At low temperatures
the presence of the wall affects the dynamics of the particles even in
the center of the tube. The curves for all temperatures show an apparent
divergence in the vicinity of the wall.  The given statements above hold 
for A and B particles and also for different wave vectors.

Based on these observations we found an empirical law which is able to describe
the $\rho$-dependence of the relaxation times of the intermediate scattering 
function. The data from Fig.~4 is described well by the functional form 
\begin{equation}
\tau_q(T,\rho)=f_q(T) \exp \left[ \Delta_q/(\rho_p - \rho) \right] ,
\label{eq_law}
\end{equation}
at least in the vicinity of 
the wall, i.e. for $\rho$$\ge$3.5, which corresponds to half of the particles.
In Eq.~(\ref{eq_law}), the penetration radius $\rho_p$ is determined  as 
mentioned above from the static properties and depends only weakly on 
temperature. The quantity $\Delta_q$ depends on particle type and the value of $q$.
The only temperature dependent quantity in this fit is the amplitude $f_q(T)$,
which also depends on particle type and $q$.
If we assume that in the supercooled state Eq.~(\ref{eq_law}) holds for most 
of the particles, the slowing down of the system in the supercooled state is 
mainly characterized by the temperature 
dependence of the amplitude $f_q(T)$. More details on this will be discussed 
elsewhere~[10].
\vspace{0.75cm}

\noindent {\bf 4. SUMMARY}
\vspace{12pt}

\noindent
We have presented the results of a computer simulation of a simple
glass former in a narrow tube.  We find the relaxation times of the
intermediate scattering function to be strongly dependent on the distance
from the wall.  We are able to describe this behavior by an empirical law
predicting a divergence at $\rho$=$\rho_p$, where $\rho_p$ is determined
from static quantities. 
We see a gradual slowing down of the dynamics of the whole system, especially 
no immobile layer close to the wall, in agreement with experiments by 
Richert [11]. Note that we expect a similar slowing down of
the dynamics in the vicinity of a wall also for other tube radii and
even in a slit geometry, and that in the case of a narrow confinement,
such as the one we investigated here, this slowing down will dominate
the dynamics of the whole system.

This work was supported by {\it Deutsche Forschungsgemeinschaft} under SFB 262/D1
and the {\it NIC} in J\"ulich.\vspace{0.75cm}

\begin{flushleft}
\noindent {\bf References}
\vspace{12pt}

\hspace{1.08mm}
[1] See, e.g., {\it Proceedings of Third Intern.~Discussion
Meeting}, J.~Non-Crysl.~Solids {\bf 235-237} (1998). \\
\hspace{1.08mm}
[2] W.~Kob, J.~Phys.: Condens. Matter {\bf 11}, R85 (1999). \\
\hspace{1.08mm}
[3] P.~Pissis, A.~Kyritsis, D.~Daoukaki, G.~Barut, R.~Pelster, and G.~Nimtz,
J.~Phys.: Condens. Matter {\bf 10},\\
\hspace{8mm} 6205 (1998).\\
\hspace{1.08mm}
[4] C.~L.~Jackson and G.~B.~McKenna, Chem.~Mater.~{\bf 8}, 2128 (1996). \\
\hspace{1.08mm}
[5] J.~Sch\"uller, B.~Yu, B.~Mel'nichenko, R.~Richert, and E.~W.~Fischer, 
Phys.~ Rev.~Lett.~{\bf 73}, 2224 (1994). \\
\hspace{1.08mm}
[6] W.~E.~Wallace, J.~H.~van Zanten, and W.~L.~Wu,~Phys.~Rev.~E, {\bf 52}, R3329 (1995).\\
\hspace{1.08mm}
[7] J.~Baschnagel and K.~Binder, J.~Phys.~I France {\bf 6}, 1271 (1996) \\
\hspace{1.08mm}
[8] Z.~T.~N\'emeth and H.~L\"owen, J.~Phys.: Condens. Matter {\bf 10}, 6189 (1998).\\
\hspace{1.08mm}
[9]  W.~Kob and H.~C.~Andersen, Phys.~Rev.~E {\bf 51}, 4626 (1995); 
Phys.~Rev.~E {\bf 52}, 4134 (1995). \\
\noindent
[10] P.~Scheidler, W.~Kob, and K.~Binder, to be published \\
\noindent
[11] R.~Richert, Phys.~Rev.~E {\bf 54}, 15762 (1996) 
\end{flushleft}

\end{document}